\documentclass[prb,twocolumn,superscriptaddress,floatfix,shortbibliography]{revtex4-2}
\usepackage{graphicx}
\usepackage{bm}
\usepackage{color,soul}
\usepackage{amsmath,amssymb,amsfonts,float,graphics,epsfig,epstopdf,color,verbatim,tabularx,bm,multirow,appendix,hyperref}

\begin{document}
\title{Strong electron-phonon coupling in Ba$_{1-x}$Sr$_x$Ni$_2$As$_2$ }
\author{Linxing Song}
\affiliation{Beijing National Laboratory for Condensed Matter Physics, Institute of Physics, Chinese Academy of Sciences, Beijing 100190, China}
\affiliation{School of Physical Sciences, University of Chinese Academy of Sciences, Beijing 100190, China}
\author{Jianguo Si}
\affiliation{Songshan Lake Materials Laboratory , Dongguan, Guangdong 523808, China}
\author{Tom Fennell}
\affiliation{Laboratory for Neutron Scattering and Imaging, Paul Scherrer Institute, CH-5232 Villigen-PSI, Switzerland}
\author{Uwe Stuhr}
\affiliation{Laboratory for Neutron Scattering and Imaging, Paul Scherrer Institute, CH-5232 Villigen-PSI, Switzerland}
\author{Guochu Deng}
\affiliation{Australian Centre for Neutron Scattering, Australian Nuclear Science and Technology Organisation, Lucas Heights, NSW 2234, Australia}
\author{Jinchen Wang}
\affiliation{Laboratory for Neutron Scattering and Beijing Key Laboratory of Optoelectronic Functional Materials and MicroNano Devices, Department of Physics, Renmin University of China, Beijing 100872, China}
\author{Juanjuan Liu}
\affiliation{Laboratory for Neutron Scattering and Beijing Key Laboratory of Optoelectronic Functional Materials and MicroNano Devices, Department of Physics, Renmin University of China, Beijing 100872, China}	
\author{Lijie Hao}
\affiliation{China Institute of Atomic Energy, PO Box 275-30, Beijing 102413, China}
\author{Huiqian Luo}
\affiliation{Beijing National Laboratory for Condensed Matter Physics, Institute of Physics, Chinese Academy of Sciences, Beijing 100190, China}
\affiliation{Songshan Lake Materials Laboratory , Dongguan, Guangdong 523808, China}
\author{Miao Liu}
\affiliation{Beijing National Laboratory for Condensed Matter Physics, Institute of Physics, Chinese Academy of Sciences, Beijing 100190, China}
\affiliation{Center of Materials Science and Optoelectronics Engineering, University of Chinese Academy of Sciences, Beijing 100049, China}
\affiliation{Songshan Lake Materials Laboratory , Dongguan, Guangdong 523808, China}
\author{Sheng Meng}
\email{smeng@iphy.ac.cn}
\affiliation{Beijing National Laboratory for Condensed Matter Physics, Institute of Physics, Chinese Academy of Sciences, Beijing 100190, China}
\affiliation{School of Physical Sciences, University of Chinese Academy of Sciences, Beijing 100190, China}
\affiliation{Songshan Lake Materials Laboratory , Dongguan, Guangdong 523808, China}
\author{Shiliang Li}
\email{slli@iphy.ac.cn}
\affiliation{Beijing National Laboratory for Condensed Matter Physics, Institute of Physics, Chinese Academy of Sciences, Beijing 100190, China}
\affiliation{School of Physical Sciences, University of Chinese Academy of Sciences, Beijing 100190, China}
\affiliation{Songshan Lake Materials Laboratory , Dongguan, Guangdong 523808, China}
\begin{abstract}
The charge density wave (CDW) or nematicity has been found to coexist with superconductivity in many systems. Thus, it is interesting that the superconducting transition temperature $T_c$ in the doped BaNi$_2$As$_2$ system can be enhanced up to six times as the CDW or nematicity in the undoped compound is suppressed. Here we show that the transverse acoustic phonons of Ba$_{1-x}$Sr$_x$Ni$_2$As$_2$ are strongly damped in a wide doping range and over the whole $Q$ range, which excludes its origin from either CDW or nematicity. The damping of TA phonons can be understood as large electron-phonon coupling and possible strong hybridization between acoustic and optical phonons as shown by the first-principle calculations. The superconductivity can be quantitatively reproduced by the change of the electron-phonon coupling constant calculated by the McMillan equation in the BCS framework, which suggests that no quantum fluctuations of any order is needed to promote the superconductivity. On the contrary, the change of $T_c$ in this system should be understood as the sixfold suppression of superconductivity in undoped compounds.
\end{abstract}


\maketitle

\section{introduction}

The electronic nematic phase breaks the rotational symmetry of the underlying lattice and in theory, its quantum fluctuations may enhance superconductivity \cite{MetlitskiMA15,LedererS15,LedererS17,KleinA19,FernandesRM19}. Experimentally, nematicity has been widely found in cuprates and iron-based superconductors, where nematic quantum fluctuations have been suggested to play important roles in promoting superconductivity \cite{FernandesRM14,ChuJH12,GallaisY13,BohmerAE14,KuoHH16,LiuZ16,GuY17,VojtaM09,HinkovV08,DaouR10,SatoY17,MurayamaH19,AuvrayN19,IshidaK20}. In the mean time, doubts on the effects of nematic fluctuations on superconductivity have also been raised \cite{HosoiS16,ChoiniereO15,ChoiniereO17, XieT22}. One of the major obstacles in further investigating the relationship between the nematicity and superconductivity in both systems is the existence of the antiferromagnetic (AFM) order and its strong fluctuations \cite{SeamusJC13,FradkinE15,KivelsonSA98,LeePA06,DaiP15}. It has been shown recently that the Ba$_{1-x}$Sr$_x$Ni$_2$As$_2$ (BSNA) system may offer an opportunity to study the role of nematic fluctuations on superconductivity without the influence from AFM fluctuations \cite{EckbergC20}.

The structures of both BaNi$_2$As$_2$ and SrNi$_2$As$_2$ are ThCr$_2$Si$_2$ type at room temperature \cite{RonningF08,SefatAS09,BauerED08}. The former changes to the triclinic structure at $T_s$ = 135 K \cite{RonningF08,SefatAS09,KothapalliK10}, while the latter stays in the tetragonal structure down to the lowest temperature. Interestingly, both of them show superconductivity at similar temperatures between 0.6 and 0.7 K despite their structural difference at low temperatures \cite{RonningF08,SefatAS09,BauerED08}. Moreover, both of them show no magnetism, which suggests that the superconducting mechanism may be associated with conventional phonon-mediated pairing \cite{SubediA08}. The structure transition in BaNi$_2$As$_2$ can be suppressed by chemical doping, such as phosphorus, copper, cobalt, and strontium, and the $T_c$ shows up to sixfold enhancement near the structural instability \cite{KudoK12,KudoK17,EckbergC18,EckbergC20,NarayanDM23}. Such a phenomenon has also been observed in other nickel pnictides \cite{HiraiD12,HlukhyyV17}, suggesting a general mechanism to enhance superconductivity.

\begin{figure}[tbp]
\includegraphics[width=\columnwidth]{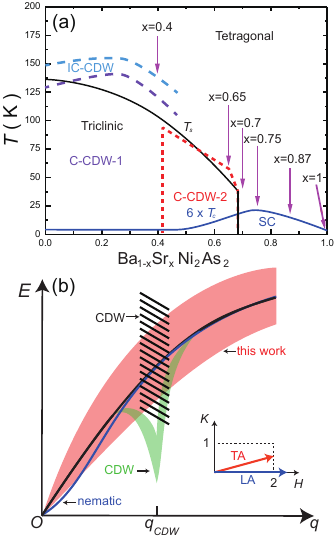}
 \caption{(a) Schematic phase diagram of Ba$_{1-x}$Sr$_x$Ni$_2$As$_2$ based on the data in Refs. \cite{EckbergC20,LeeS21}. SC stands for superconductivity. The arrows indicate the samples measured by INS in this work. (b) Schematic effects of nematic fluctuations and CDW on the acoustic phonons. The black and blue lines are unaffected phonon dispersion and that under nematic fluctuations, respectively. The change of phonons by CDW are illusrated by the green shaded area and black oblique lines, respectively. The pink area indicates the results of this work. The inset shows the schematic scans to measure the TA and LA phonons around the (2,0,0).
}
\label{fig1}
\end{figure}

The enhancement of superconductivity in doped BaNi$_2$As$_2$ systems may be associated with the quantum fluctuations of two phases. The first one is nematicity, which has been suggested to exist in both BSNA and BaNi$_2$(As,P)$_2$ systems \cite{EckbergC20,MerzM21,YaoY22}. The nematic transition temperature extracted from the Curie-Weiss-like fitting of the elastoresistivity becomes zero near optimal doping, which resembles the case in iron-based superconductors \cite{ChuJH12,KuoHH16,LiuZ16,GuY17}. The second phase involved is the charge density wave (CDW). Two types of CDWs have been found in BaNi$_2$As$_2$, an incommensurate CDW (IC-CDW) above $T_s$ and a commensurate CDW (C-CDW-1) below it \cite{LeeS19}. In BSNA, a new commensurate CDW (C-CDW-2) appears above $x$ = 0.4 \cite{LeeS21}, as shown in Fig. \ref{fig1}(a). The appearance and enhancement of superconductivity near the CDW instability have been widely found in many other systems \cite{MorosanE06,SiposB08,WagnerKE08,MorosanE10,LiuY16}. Whether the BSNA system falls into the same category is interesting and unclear. We note that both CDW and nematic order have effects on phonon spectra, as roughly illustrated in \ref{fig1}(b), and thermodynamical measurements have indicated significant phonon softening when the CDW disappears \cite{KudoK12,KudoK17,EckbergC20,MeingastC22}. Thus, it is crucial to further investigate the pairing mechanism by studying the phonons. 

In this work, we studied the acoustic phonons in BSNA by the inelastic neutron scattering (INS) technique. Wide Q-range damping of the transverse acoustic (TA) phonons are found in all samples around the opitmal doping. Such large damping of TA phonons can be understood by neither CDW nor nematic fluctuations. By comparing with the first-principle calculations, we show that the superconductivity around the optimal doping level can be well understood by the McMillan equation in the BCS theory without invoking the quantum fluctuations from either CDW or nematicity.

\section{experiments}

Single crystals of BSNA were grown by the self-flux method as reported previously \cite{EckbergC20}. The INS experiments were carried out on the thermal triple-axis spectrometers EIGER \cite{StuhrU17} at SINQ, Switzerland, TAIPAN at ANSTO, Australia \cite{DanilkinSA07}, and cold neutron multiplexing spectrometer BOYA at CARR, China. The information of all samples is listed in the Supplemental Material \cite{supp}. The samples were measured in the [$H$,$K$,0] scattering plane with the momentum transfer $\textbf{Q}$ = $H\textbf{a}^*$+$K\textbf{b}^*$+$L\textbf{c}^*$, where the reciprocal lattice parameters $\textbf{a}^*$, $\textbf{b}^*$ and $\textbf{c}^*$ are defined in the tetragonal lattice with $a = b \approx$ 4.144 \AA and $c \approx$ 11.633 \AA. The TA and longitudinal acoustic (LA) phonons were measured by scans along [2,$K$,0] and [$H$,0,0], respectively, as shown in the inset of Fig. \ref{fig1}(b). The resolutions of the instruments were calculated by the ResLibCal program \cite{FarhiE13}, where the effects of sample mosaics have been included. The first-principles calculations were preformed via the QUANTUM ESPRESSO package \cite{GiannozziP21} with the virtual crystal approximation method \cite{BellaicheL00}. The ultrasoft pseudopotentials with the generalized gradient approximation was used to create the pseudopotential of fake atoms, and which is parametrized by the Perdew-Burke-Ernzerhof function \cite{PerdewJP96}. Phonon dispersion curves were calculated based on the density functional perturbation theory \cite{BaroniS01}, where a denser 16$\times$16$\times$16 (16$\times$16$\times$12) $k$-point grid and a 4$\times$4$\times$4 (4$\times$4$\times$3) $q$-point grid were employed for the electron-phonon coupling (EPC) calculations of the tetragonal (triclinic) phase.

\begin{figure}[tbp]
\includegraphics[width=\columnwidth]{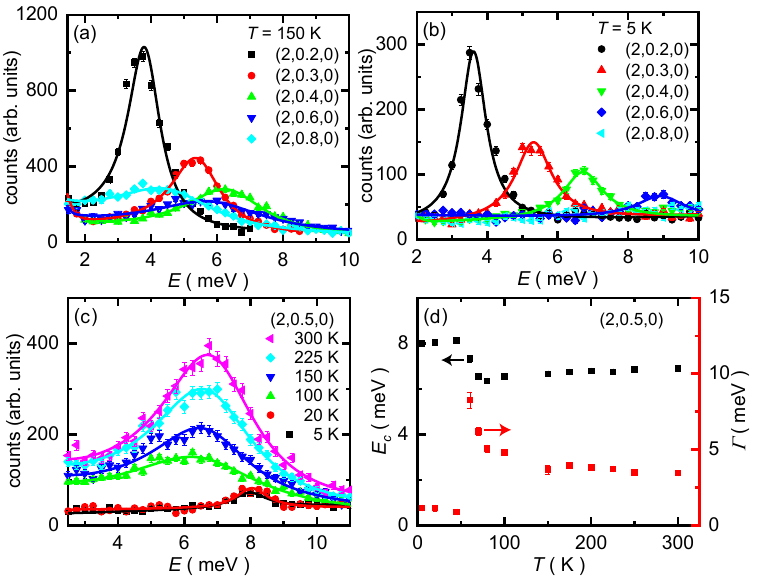}
 \caption{The results of TA phonons for the $x$ = 0.65 sample. (a), (b) Constant-$Q$ scans along [2,$K$,0]  at 150 and 5 K, respectively. The solid lines are fitted by Eq. (\ref{damp}). (c) Constant-$Q$ scans at (2,0.5,0) at various temperatures. The solid lines are fitted by Eq. (\ref{damp}). (d) The temperature dependence of phonon energy and width at (2,0.5,0).
}
\label{fig2}
\end{figure}

\section{results and discussions}

Figures \ref{fig2}(a) and \ref{fig2}(b) show constant-$Q$ scans at [2, $K$, 0] for the TA phonons in the $x$ = 0.65 sample at 150 and 5 K, respectively. Since the crystal structure at 5 K is triclinic, the nominal [2, $K$, 0] in the tetragonal notation approximately corresponds to [2, $K$, -1-$K$/2] in the triclinic notation. To quantitatively study the phonon dispersion, we use the following damped harmonic-oscillator function \cite{ShiraneG02} to fit the constant-$Q$ scans,
\begin{equation}
S(Q,E) = \frac{A}{E_c(1 - e^{-E/k_BT})}\frac{\Gamma/2}{(E-E_c)^2+\Gamma^2/4}
\label{damp}
\end{equation}
\noindent where $E_c$, $\Gamma$ and $A$ are the phonon energy, the peak full-width at half-maximum (FWHM) and the fitting constant. The backgrounds are fitted by a constant plus a Gaussian function with the center fixed at zero energy to account for the tail from the elastic incoherent nuclear scattering. 

Figure \ref{fig2}(c) shows the constant-$Q$ scans at (2,0.5,0) at various temperatures, which is the $Q$ position associated with the C-CDW-2. Figure \ref{fig2}(d) gives the temperature dependence of the corresponding $E_c$ and $\Gamma$. The value of $E_c$ only slightly decreases with decreasing temperature when approaching $T_s$, whereas $\Gamma$ shows a quick upturn below about 80 K. While such upturn is associated with the CDW transition, the almost temperature-independent large $\Gamma$ above 100 K should not be. Below the structural or CDW transition, $E_c$ changes because of the triclinic structure at low temperatures and $\Gamma$ becomes much smaller.

\begin{figure}[tbp]
\includegraphics[width=\columnwidth]{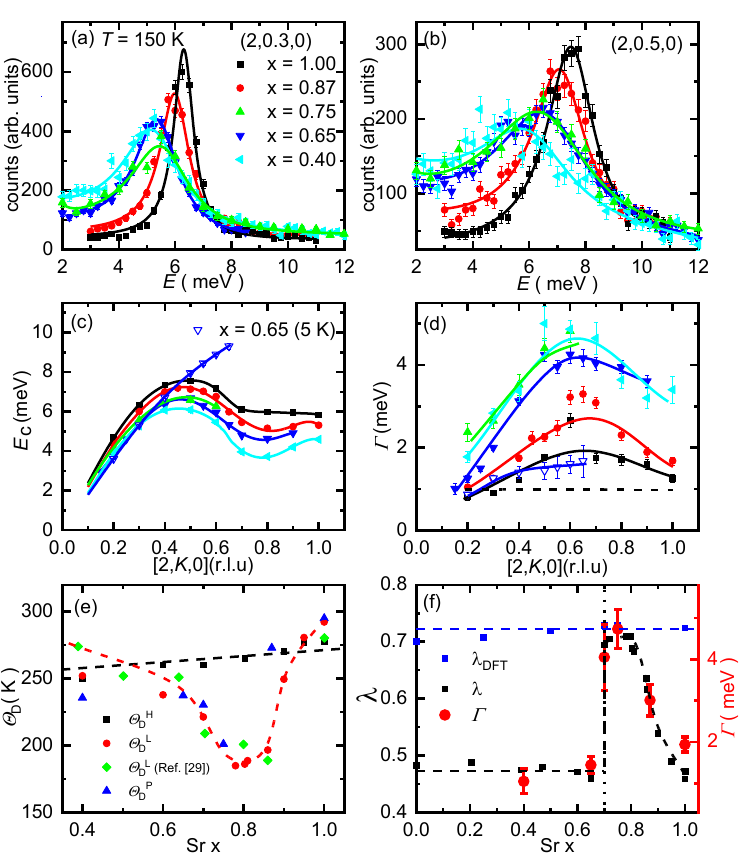}
 \caption{Summary of the TA phonons in BSNA. (a), (b) Constant-$Q$ scans at (2,0.3,0) and (2,0.5,0), respectively, at 150 K. The solid lines are fitted by Eq. (\ref{damp}). (c) The phonon dispersions along [2,$K$,0] of all samples. The solid lines are guides for the eye.  (d) The fitted $\Gamma$ along  [2,$K$,0]. The solid lines are guides for the eye. The dashed line is $\Gamma_{cal}$ as described in the main text. We note that the data here were all obtained on EIGER with the same configuration, so only one $\Gamma_{cal}$ is considered. The large $\Gamma$ value at (2,0.6,0) for the $x$ =1 sample is due to the resolution effect \cite{supp}.  (e) The doping dependence of the Debye temperatures obtained by various methods as described in the main text. $\Theta_D^H$, $\Theta_D^L$, and $\Theta_D^P$ are the Debye temperatures calculated based on high-temperature specific heats, low-temperature specific heats, and the phonon velocities, respectively. The dashed lines are guides to the eye. (f) The doping dependence of EPC constant $\lambda_{DFT}$  and $\lambda_{EXP}$ obtained by LDA calculations and Eq. (\ref{lambda}), respectively. The phonon energy width $\Gamma$ at (2,0.5,0) is also shown. The dashed lines are guides to the eye. 
}
\label{fig3}
\end{figure}

Figures \ref{fig3}(a) and \ref{fig3}(b) show the constant-$Q$ scans at (2,0.3,0) and (2,0.5,0), respectively, for other samples at 150 K. The data at 5 K of the $x >$ 0.75 samples show no significant difference as there is no structural transition \cite{supp}. With decreasing $x$, the peak position decreases while the width becomes broader. The data can all be well fitted by Eq. (\ref{damp}), which gives TA phonon dispersions and FWHMs in Figs. \ref{fig3}(c) and (d), respectively. It is clear that TA phonons become softer with decreased $x$ while the damping effect gets larger in the tetragonal structure (150 K). We note that a minimum appears around $K \approx$ 0.8, which may come from the low-energy optical phonons that are strongly coupled with the TA phonons as shown later, considering that the condition for measuring TA phonons by INS is not well satisfied anymore at large $K$'s. The phonon dispersion of the $x$ = 0.65 sample at 5 K is different from that at 150 K, which is due to different crystal structures. The values of $\Gamma$ of the $x$ = 1 sample and that of the $x$ = 0.65 sample at 5 K are close to $\Gamma_{cal}$ for undamped phonons, which is calculated by considering the phonon dispersion and instrument resolution \cite{ChesserNJ72} at small $Q$'s. Therefore, we can consider that there is no or little damping effect for the TA phonons in the $x$ = 1 sample and the TA phonons in the $x \leq 0.65$ sample with the triclinic structure.

The evolution of the TA phonons provides a quantitative understanding on the change of the Debye temperature. The Debye temperature $\Theta_D^H$ can be obtained by fitting the high-temperature specific heat with the Debye model \cite{supp}, which is shown in Fig. \ref{fig3}(e). $\Theta_D^H$ smoothly decreases with decreasing $x$, which is consistent with the lowering of the TA phonon dispersion as shown in Fig. \ref{fig3}(c). On the other hand, a previous report has shown that the Debye temperature $\Theta_D^L$ obtained by the low-temperature specific heat drops dramatically around the optimal doping level \cite{EckbergC20}, as shown together with our results in Fig. \ref{fig3}(e) \cite{supp}. We note that $\Theta_D^L$ is calculated by fitting the low-temperature specific heat $C$ with $\gamma T + \beta T^3$, where $\beta$ = 12$\pi^4k_B/5(\Theta_D^L)^3$ according to the low-temperature limit of the Debye model. As $\Theta_D^L$ is proportional to the effective sonic velocity $\nu_{eff}$, which can be estimated by the LA and TA phonon velocities ($\nu_L$ and $\nu_T$) at low energies as $\nu_{eff}^{-3}$ = (1/3)$\nu_L^{-3}$+(2/3)$\nu_T^{-3}$ \cite{supp}. By considering the broadening of the TA phonons that effectively gives a distribution of $\nu_T$, we can calculate $\Theta_D^P$ as shown in Fig. \ref{fig3}(e), which quantitatively explains the drop of $\Theta_D^L$ around the optimal doping level.

The large damping of TA phonons typically come from strong EPC. According to BCS theory, the EPC constant $\lambda$ is related to $T_c$ as the following empirical equation \cite{McMillanWL68}
\begin{equation}
\lambda = \frac{1.04+\mu^*ln(\Theta/1.45T_c)}{(1-0.62\mu^*)ln(\Theta/1.45T_c)-1.04},
\label{lambda}
\end{equation}
\noindent where $\mu^*$ is the pseudopotential. The value of $\mu^*$ may be taken from 0.1 to 0.2. Here we choose $\mu^*$ to be 0.16 to compare with the theoretical calculation of $\lambda_{DFT}$ as discussed later. According to the experimental values of $\Theta_D^L$ [Fig. \ref{fig3}(e)] and $T_c$, the doping dependence of $\lambda_{EXP}$ is shown in Fig. \ref{fig3}(f). We note that changing $\mu^*$ only changes the absolute value of $\lambda_{EXP}$ and does not change the overall doping dependence. With decrease $x$ from 1 to 0.7, $\lambda_{EXP}$ quickly increases from about 0.45 to 0.7. We note that such change of $\lambda$ is quantitatively consistent with the change of $\Gamma$ at (2,0.5,0), further suggesting the origin of the phonon broadening as from the EPC coupling. For $x \leq$ 0.65 where the low-temperature structure becomes triclinic and no damping of TA phonons presents, $\lambda_{EXP}$ drops below 0.5. This means that the change of $T_c$ in the BSNA system can be readily explained by the increase of EPC and the decrease of the Debye temperature within the mechanism of BCS theory. 

\begin{figure}[tbp]
\includegraphics[width=\columnwidth]{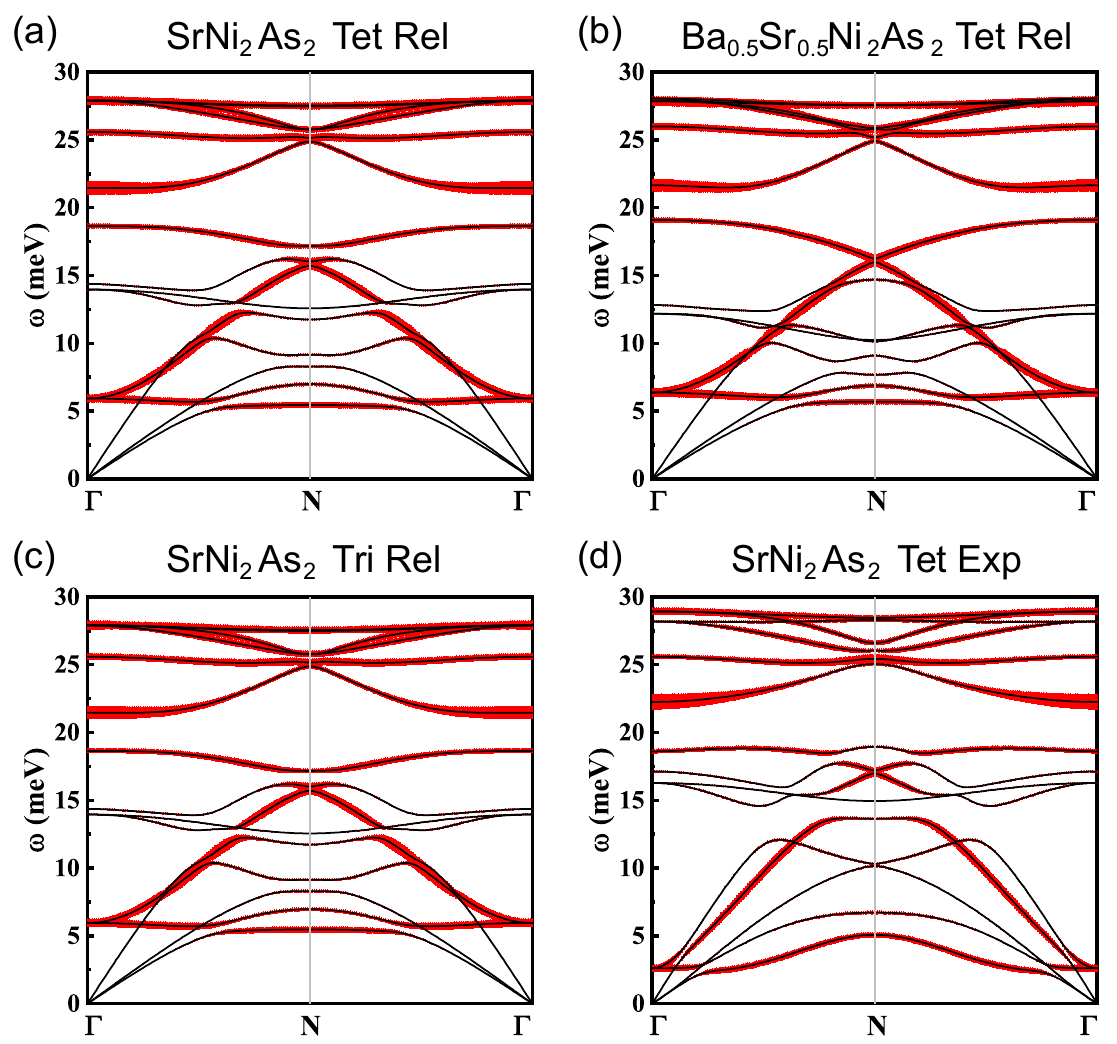}
 \caption{Phonon dispersions weighted by the phonon linewidth (red circles) for fully relaxed (Rel) (a) tetragonal (Tet) SrNi$_2$As$_2$, (b) tetragonal Ba$_{0.5}$Sr$_{0.5}$Ni$_2$As$_2$ and (c) triclinic (Tri) SrNi$_2$As$_2$. (d) Phonon dispersions weighted by the phonon linewidth calculated by considering the experimental (Exp) parameters for tetragonal SrNi$_2$As$_2$. 
}
\label{fig4}
\end{figure}

To understand the broadening of the TA phonons, we calculated the phonon spectra and EPC by the first-principles calculations. Figures \ref{fig4}(a) and \ref{fig4}(b) show the calculated phonon dispersion weighted by the phonon linewidth for tetragonal SrNi$_2$As$_2$ and Ba$_{0.5}$Sr$_{0.5}$Ni$_2$As$_2$, where the crystal structures are fully relaxed. We also create a triclinic SrNi$_2$As$_2$ by changing the Ba into Sr in the structure of triclinic BaNi$_2$As$_2$ and fully relax the lattice, whose spectra are shown in Fig. \ref{fig4}(c). It is clear that there is no substantial change for different doping levels and structures. Moreover, the EPC constant $\lambda$ also shows little doping dependence, as already shown in Fig. \ref{fig3}(f). We note that while the large $\lambda$ is consistent with the experimental value around the optimal doping level, the calculated broadening of the phonons happens mostly for optical phonons. We thus fix the structure of SrNi$_2$As$_2$ by the experimental parameters without relaxation. The acoustic phonons at lower energies become significantly broader, as shown in Fig. \ref{fig4}(d), suggesting large acoustic-optical phonon hybridization in real materials. We note that the value of $\lambda$ remains unchanged even when using the experimental lattice parameters.

Our results on the large damping of TA phonons cannot be explained by nematic fluctuations. Previous measurements on iron-based superconductors have shown that the TA phonons become slightly softened at low energies and small $q$ near Bragg peaks in the presence of nematic fluctuations \cite{WeberF18,LiY18,MerrittAM20} as shown in Fig. \ref{fig1}(b), because nematic fluctuations are long wavelength. Moreover, such softening should be reduced with increasing temperature due to the decrease of nematic susceptibility \cite{WeberF18,LiY18,MerrittAM20}. However, our results show that the TA phonons in the optimally doped samples are significantly damped over the whole $Q$ range with little temperature dependence, which clearly cannot be understood within the picture based on nematic fluctuations. It should be noted that even the presence of nematic fluctuations in nickel pnictides may be in doubt \cite{FrachetM22}. 

The large damping effects cannot come from CDW fluctuations either. Phonon softening and damping effects have been widely observed in many CDW systems \cite{CarneiroK76,WeberF11,MonceauP12,BlackburnE13,MiaoH18} and BaNi$_2$As$_2$ \cite{SouliouSM22,SongY22}. However, we note that such softening and damping of phonons are confined around the CDW wave vectors as shown in Fig. \ref{fig1}(b) and have strong temperature dependence associated with the CDW order. In our case, the damping of TA phonons is found over the whole $Q$ and temperature ranges in the optimally doped samples where no CDW order occurs. 

We attribute the large damping effects of TA phonons to the large EPC and acoustic-optical phonon hybridization in this system. As shown above, theoretical calculations have revealed strong electron phonon couplings and large phonon linewidths of many optical phonons. Indeed, the linewidth of the lowest optical branch around 6 meV (Fig. \ref{fig4}), i.e. the $E_g$ mode in Raman scattering, is already very large at room temperature in BaNi$_2$As$_2$ \cite{YaoY22}. The FWHM of the $E_g$ mode is about 2.7 meV, close to the values of our results at large $K$'s. The TA phonons are thus broadened through the acoustic-optical phonon hybridization. It has been shown in some thermoelectric materials that strong acoustic-optical hybridization can lead to low lattice thermal conductivity $\kappa_{lat}$ \cite{LiW16,BaiFX23}. In our samples, we have found that $\kappa_{lat}$ in the optimally doped samples is indeed strongly suppressed \cite{WuXW23}, consistent with the large acoustic-optical phonon hybridization.

The large EPC naturally explains the superconductivity of the BSNA system within the conventional BCS framework. As shown by the above calculations and previous studies \cite{SubediA08}, the EPC constant $\lambda$ is supposed to be about 0.7 without introducing strong electron-electron correlations. We note that recent studies have shown the weak nature of correlations in BSNA \cite{NarayanDM23}, suggesting the reliability of our calculations. Such large $\lambda$ naturally gives a $T_c$ higher than 3 K according to the McMillan equation \cite{McMillanWL68}. Therefore, the superconductivity and phonon spectra around optimal doping are very conventional without the need to invoke quantum fluctuations of either CDW or nematicity. This is contrary to the wisdom that the optimal superconductivity in this system may come from an undoped compound and the quantum fluctuations associated with the suppression of the order in this parent compound \cite{EckbergC20,NarayanDM23}. Instead, the change of superconductivity with doping should be understood as the sixfold suppression of the superconductivity in undoped compounds due to the weakening of the EPC. For BaNi$_2$As$_2$, the decrease of EPC can be easily attributed to the CDW as the change of lattice structure has little effects on the EPC. On the other hand, it is unclear what mechanism results in the suppression of EPC in SrNi$_2$As$_2$, which show no signs of CDW or other order.

\section{conclusions}

In conclusion, we have shown large damping effects of TA phonons in BSNA, which comes from large EPC and can well reproduce the optimal superconductivity within the conventional framework of the BCS theory. Neither nematicity nor CDW is necessary to be introduced in this mechanism. Our results suggest that the superconductivity in this system should not be understood as the sixfold enhancement from parent compounds but rather is sixfold suppressed in them. The origin of the large EPC and its suppression in SrNi$_2$As$_2$ need to be further studied.

\acknowledgments

S. L. thanks Prof. Pengcheng Dai for discussions. This work is supported by the National Key Research and Development Program of China (Grants No. 2022YFA1403400, No. 2021YFA1400400, No. 2021YFA1400020, No. 2023YFA1406100, No. 2018YFA0704200, and No. 2023YFA1406500), the National Natural Science Foundation of China (Grants No. 12025407, No. 12004426, and No. 12304185), the Chinese Academy of Sciences (Grants No. XDB33000000, No. GJTD-2020-01, No. CAS- WX2023SF-0101, and No. ZDBS-LY-SLH007).

L.S. and J.S. contributed equally to this work.

\end{document}